\documentclass[twocolumn,showpacs,aps,epsfig,nofootinbib,floatfix]{revtex4}

\usepackage{graphicx}
\usepackage{epstopdf}
\usepackage{latexsym}
\usepackage{amssymb}
\usepackage{amsmath}

\usepackage[center]{subfigure}

\begin{document}

 \newcommand{\bq}{\begin{equation}}
 \newcommand{\eq}{\end{equation}}
 \newcommand{\bqn}{\begin{eqnarray}}
 \newcommand{\eqn}{\end{eqnarray}}
 \newcommand{\nb}{\nonumber}
 \newcommand{\lb}{\label}
\newcommand{\PRL}{Phys. Rev. Lett.}

\title{Measurement of net electric charge and dipole moment of dust aggregates in a complex plasma }

\author{Razieh Yousefi }
\email{Raziyeh_Yousefi@baylor.edu}

\author{Allen B. Davis}

\author{Jorge Carmona-Reyes}
\email{Jorge_Carmona_Reyes@baylor.edu}

\author{Lorin S. Matthews}
\email{Lorin_Matthews@baylor.edu}

\author{Truell W. Hyde}
\email{Truell_Hyde@baylor.edu}

\affiliation{ CASPER, Physics Department, Baylor
University, Waco, TX 76798-7316, USA\\
}
\date{\today}

\begin{abstract}

Understanding the agglomeration of dust particles in complex plasmas requires a knowledge of the basic properties such as the net electrostatic charge and dipole moment of the dust. 
In this study, dust aggregates are formed 
from gold coated mono-disperse spherical melamine-formaldehyde monomers in a radio-frequency (rf) argon discharge plasma. The behavior of observed dust aggregates is 
analyzed both by studying the particle trajectories and by employing computer models examining 3D structures of aggregates and their interactions and rotations as induced by torques arising from their dipole moments. These 
allow the basic characteristics of the dust aggregates, such as the electrostatic charge and dipole moment, to be determined. It is shown that the experimental results support the predicted values from computer models for aggregates in these environments.

\end{abstract}

\pacs{52.27.Lw; 52.58.Qv; 52.70.Nc; 52.70.Ds}

\maketitle

\section{INTRODUCTION}
\renewcommand{\theequation}{1.\arabic{equation}} \setcounter{equation}{0}

The agglomeration of micron and sub-micron sized particles is a fundamental process that occurs over a range of environments.  Dust plays a key role in the
 evolution of molecular clouds and the early stage of star and planet formation. The processes governing the evolution of the system are tied to the properties of the dust grains, such
 as the size distribution and porosity (fluffiness) of the aggregates \cite{1a,1b}.  The formation of ice particles on dust grains in Earth’'s upper atmosphere gives rise
 to unusual phenomena such as noctilucent clouds and localized electron depletions (``biteouts'') associated with these clouds \cite{1c}.
 These ice particles also reflect and absorb energy, influencing atmospheric dynamics \cite{1d}. 
 Dust is also readily found in magnetic confinement fusion devices, which has consequences for the safety of fusion reactors as well as
 their operation and performance \cite{1e}.  Additionally the coagulation of dust in plasma, following gas-phase nucleation, is a source of
 contamination in semiconductor processing \cite{1f,1g}.

 In many of the environments mentioned above 
the dust is charged, either by being immersed in a radiative or plasma environment
 or through triboelectric charging. While the microphysics of the coagulation process is
 fundamental to all of these areas, it is not yet well understood. One of the reasons for this is due to the fact that a knowledge of the basic properties of the dust particles is  required to better understand and 
explain these phenomena. The particle electrostatic charge and dipole moment 
are two of the most important parameters which help define the dusty plasma since they determine both the particle interactions with themselves as well as with the plasma 
particles and existing electromagnetic fields. 

Grains in a plasma environment often attain charges sufficiently large that they 
are equally affected by local electrical and gravitational fields \cite{4a,4b}.
One method for calculating the charge of isolated dust grains in a Maxwellian plasma is orbital-motion limited (OML) theory, which has been applied to spherical dust
 grains \cite{4c,4d}, cylindrical dust particles \cite{4e} and less symmetric isolated particles using a non-spherical probe \cite{4f}. The charge on non-isolated dust grains, where interactions between dust particles is taken into account, has also been investigated theoretically \cite{4h}. 
Measurement of the charge on dust grains in laboratory experiments has also been performed for  both isolated \cite{4i} and non-isolated dust grains \cite{4j}. For most conditions experimental results agree fairly well with theoretical predictions.

Dust grains in astrophysical and laboratory plasmas are not necessarily spherically symmetric; the grains
 are typically elongated \cite{4k, 4l} or aggregates consisting of many small subunits \cite{4ka}. It has been shown that aggregates in both laboratory and astrophysical environments tend to acquire more charge compared to spherical
 grains of the same mass due to their porous structure \cite{4la,4lb,4lc,8b}. 
 A fluffy aggregate, consisting of many spherical monomers, has charge distributed over its irregular surface which leads to a nonzero dipole moment.

\begin{figure}
\includegraphics[width=.3615\textwidth]{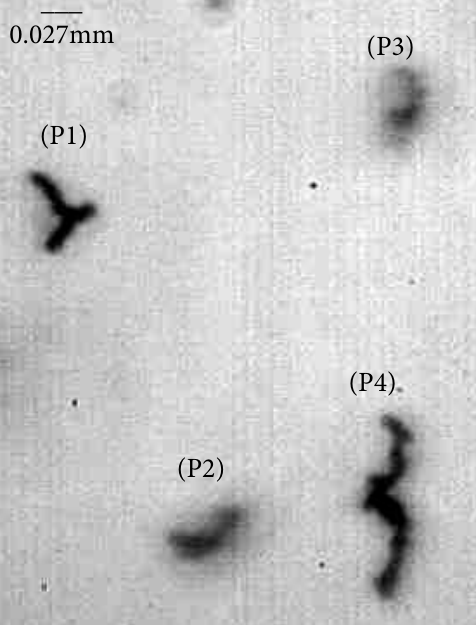}
\caption{A representative single frame of rotating aggregates showing four particles. Rotation of $P1 -P3$ about the vertical axis was observed,
 while $P4$ was moving from right to left without any rotation, probably due to its small horizontal electric dipole moment.  }
\label{fig1}

\end{figure}

The electric charge and dipole moment of aggregates play a very important role in the coagulation process. An opposite charge polarity can lead to enhanced coagulation as a result of mutual electrostatic attraction \cite{1c}.
However grains which are in a plasma environment usually obtain the same charge polarity. The coagulation of like-charged grains, while less efficient than the coagulation of oppositely-charged grains, may be enhanced if dipolar mechanisms \cite{8ba} such as charge-dipole
 interactions \cite{4q,4r}, which have been observed experimentally \cite{4n} and modeled numerically \cite{4o} or dipole-dipole interactions \cite{4m,4ma}, are taken into account.

It is evident that any proper explanation of agglomeration of dust particles requires a fundamental understanding of the electrostatic charge and dipole moment of dust aggregates.  
 The electrostatic charge
 and dipole moment of individual particles in a cloud of spherical dielectric particles, dispersed in air passing through a  non-uniform electric field,
 has been experimentally measured in \cite{8c} by studying the particles' trajectories.
In numerical studies, the electric charge and dipole moment of aggregates in a plasma environment has been calculated using modified $OML$
 theory \cite{13} and the interaction between two charged grains modeled by calculating the torques and accelerations due to the charged aggregates or external electric fields \cite{8a,8b}. However a direct study of both
 electrostatic charge and dipole of dust particles in laboratory plasmas has not yet been made.
 In this study, we present a model for these complex plasma environments, where dust aggregates formed 
in a laboratory plasma and their basic properties are investigated. These results are then compared with an existing numerical model.

 \section{EXPERIMENT}

\begin{figure}[htb]
\centering
  \begin{tabular}{@{}cccc@{}}
    \includegraphics[width=.10\textwidth]{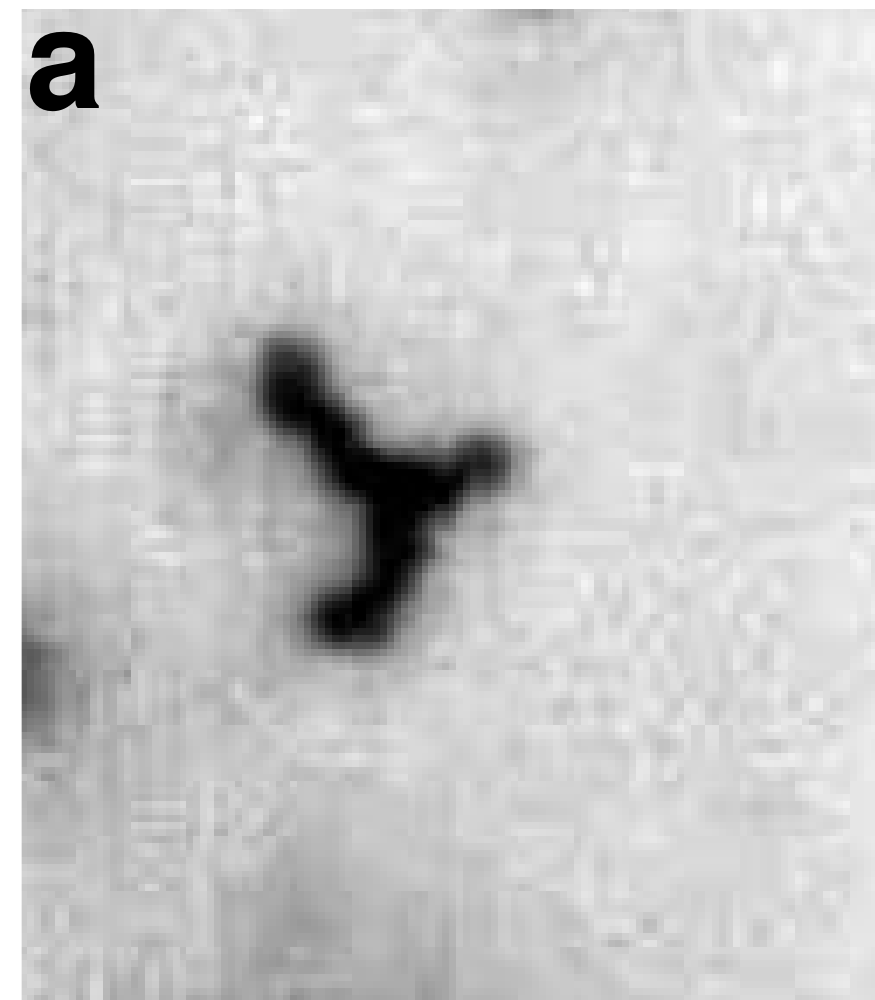} &
    \includegraphics[width=.10\textwidth]{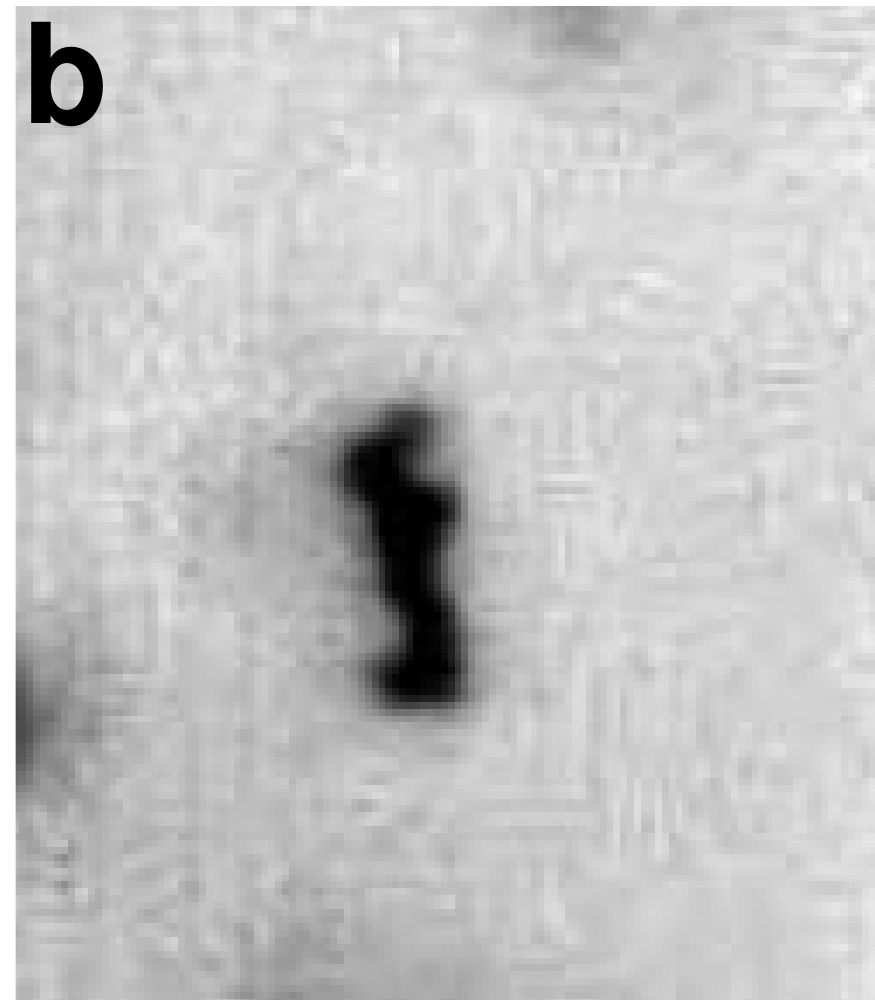} &
    \includegraphics[width=.10\textwidth]{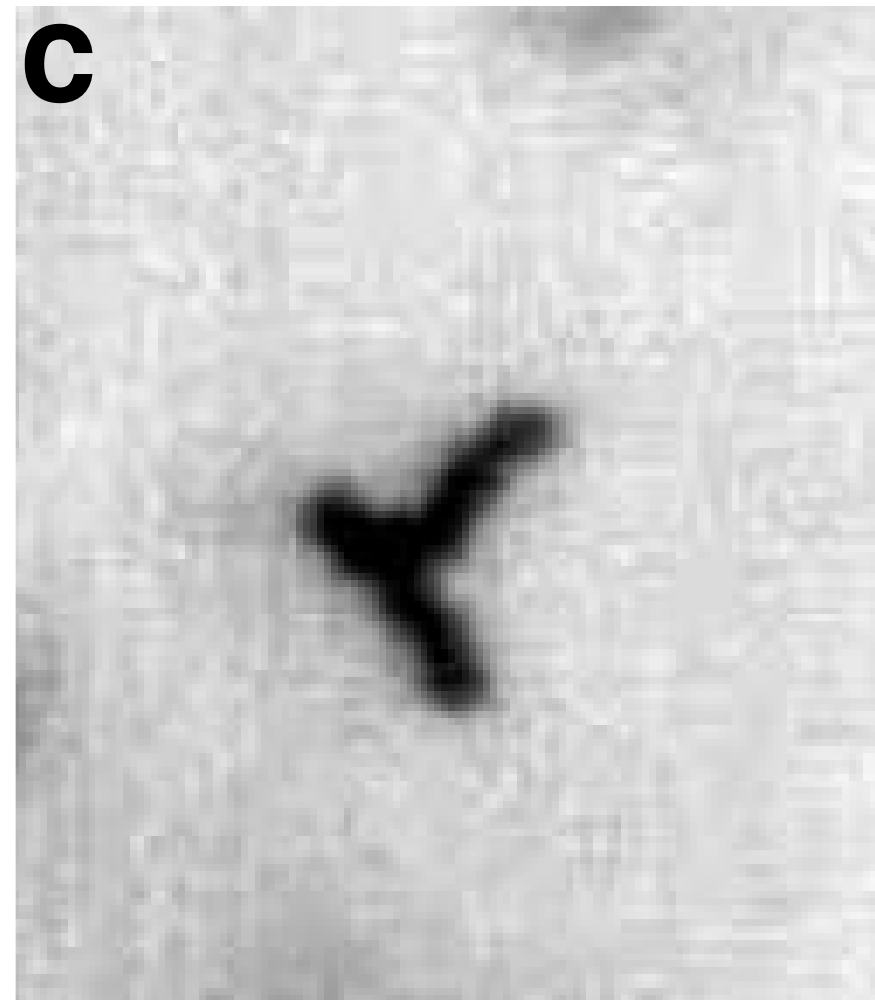} &
    \includegraphics[width=.10\textwidth]{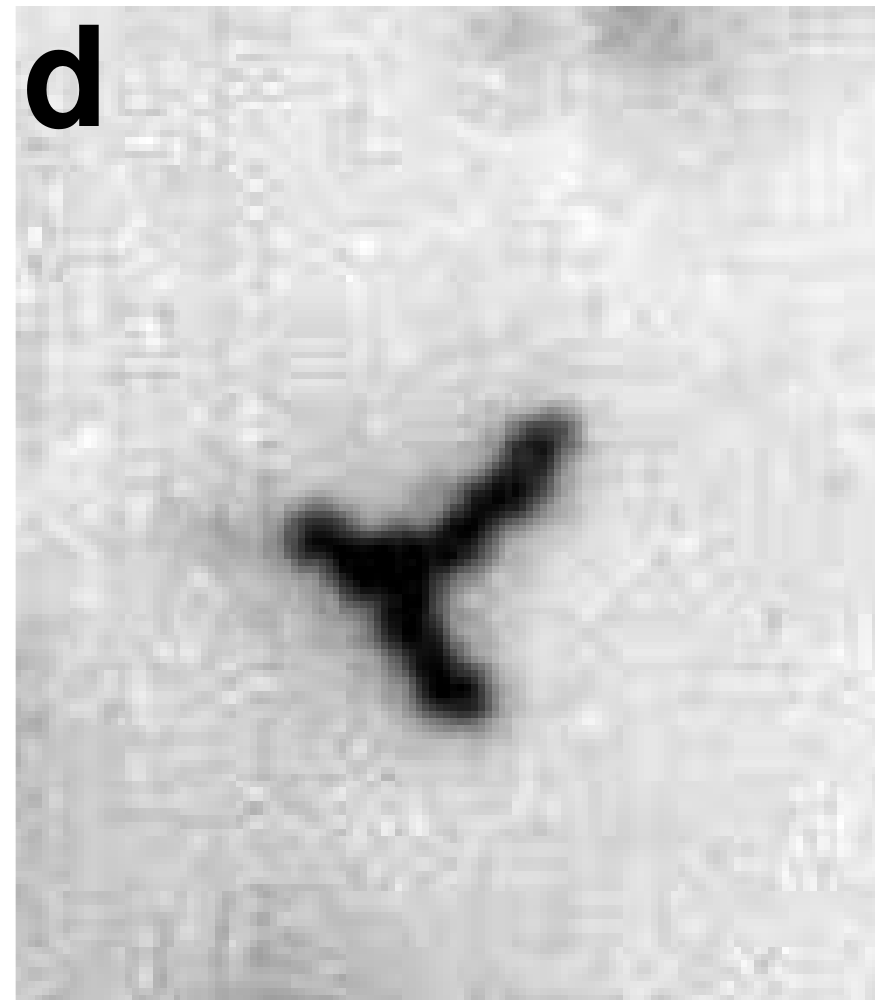}   \\
    \includegraphics[width=.10\textwidth]{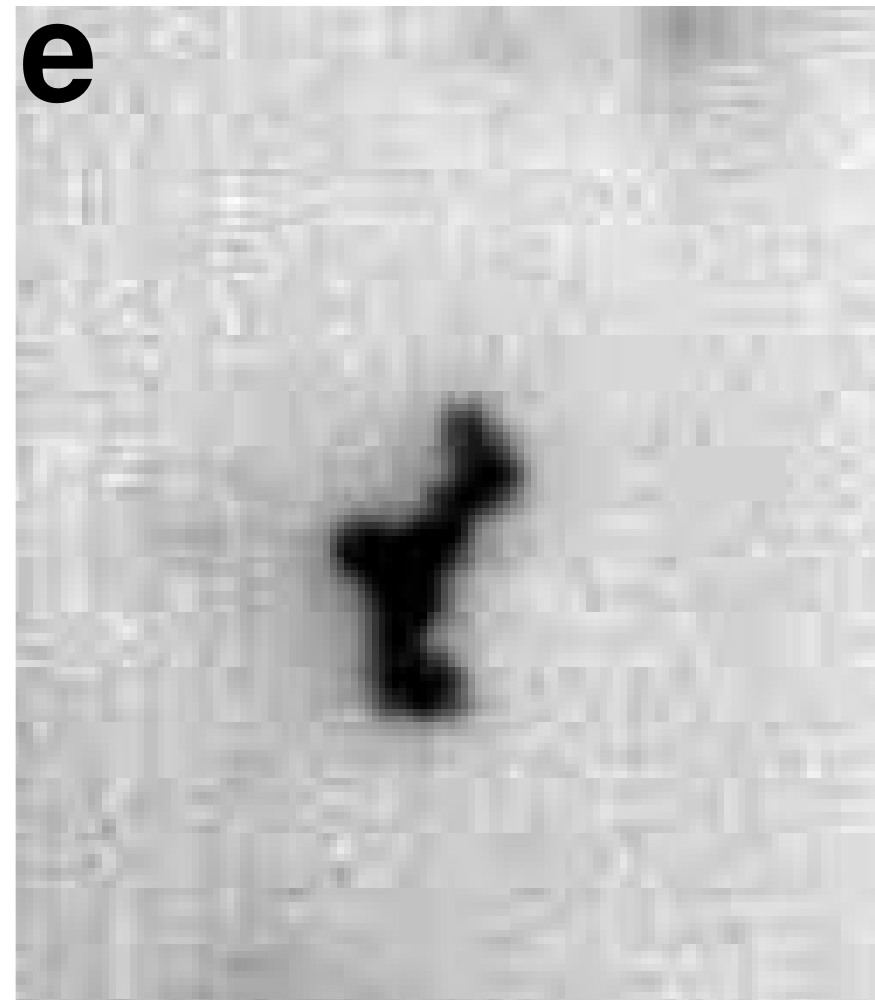} &
    \includegraphics[width=.10\textwidth]{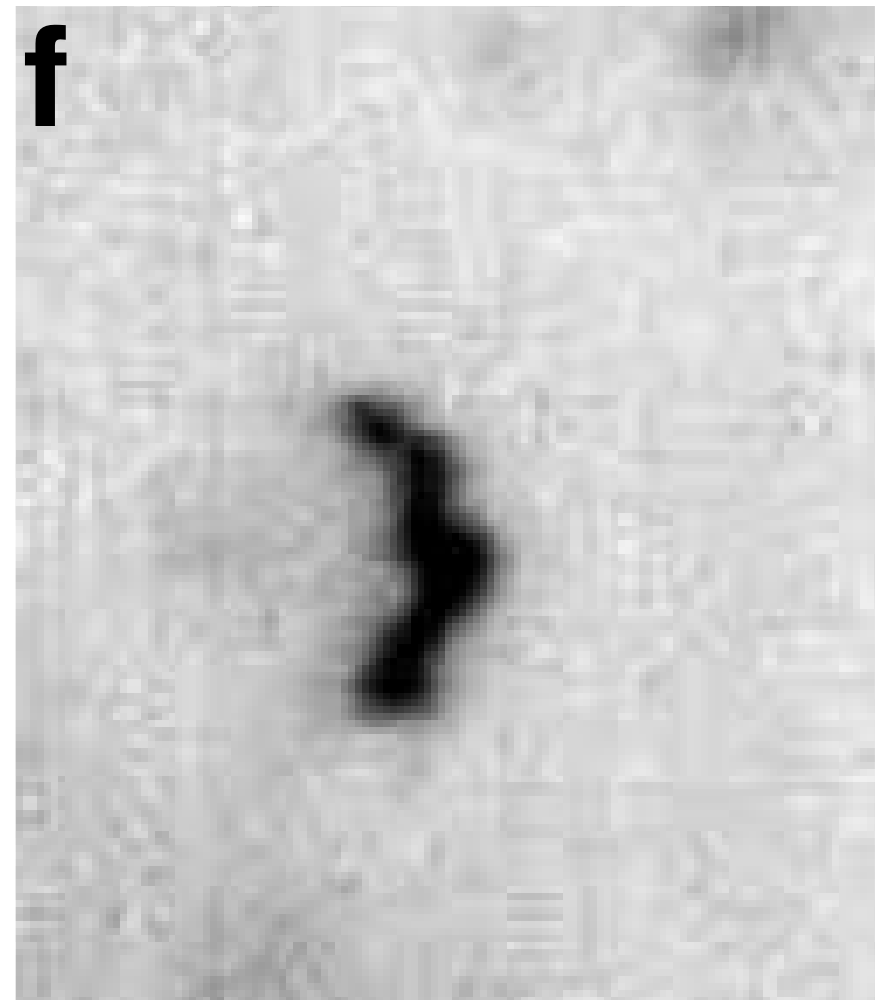} &
    \includegraphics[width=.10\textwidth]{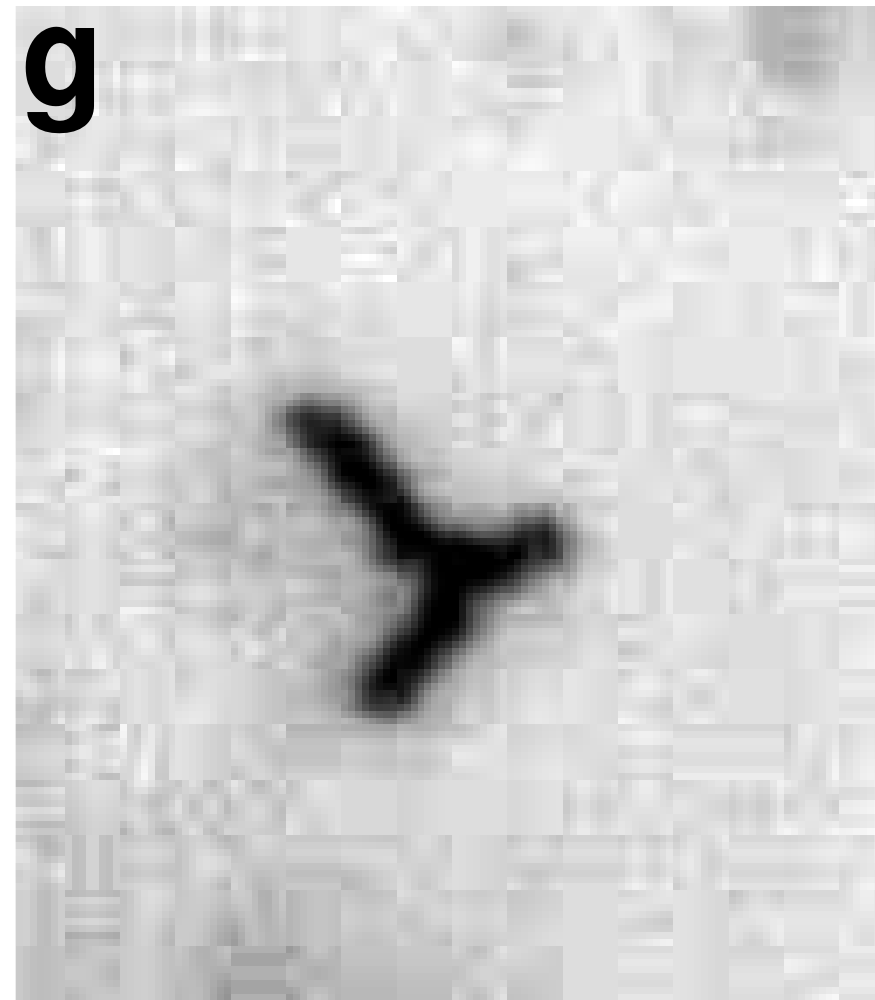} &
    \includegraphics[width=.10\textwidth]{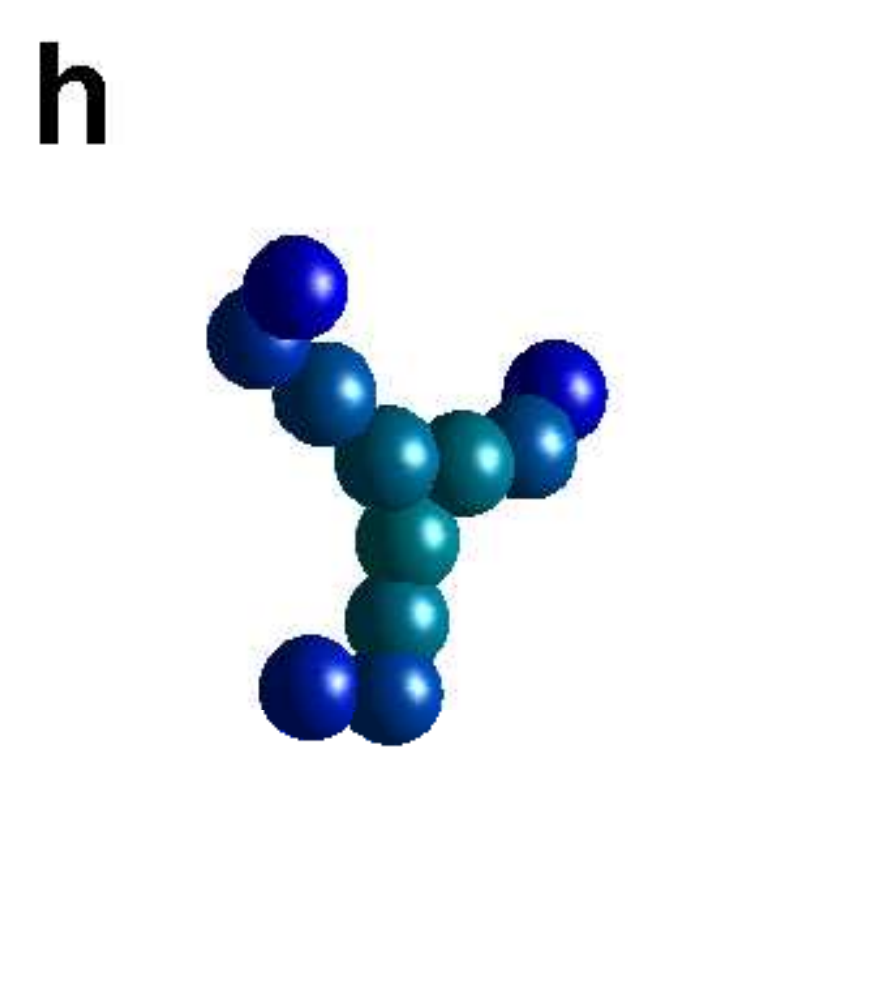}   \\
    \includegraphics[width=.10\textwidth]{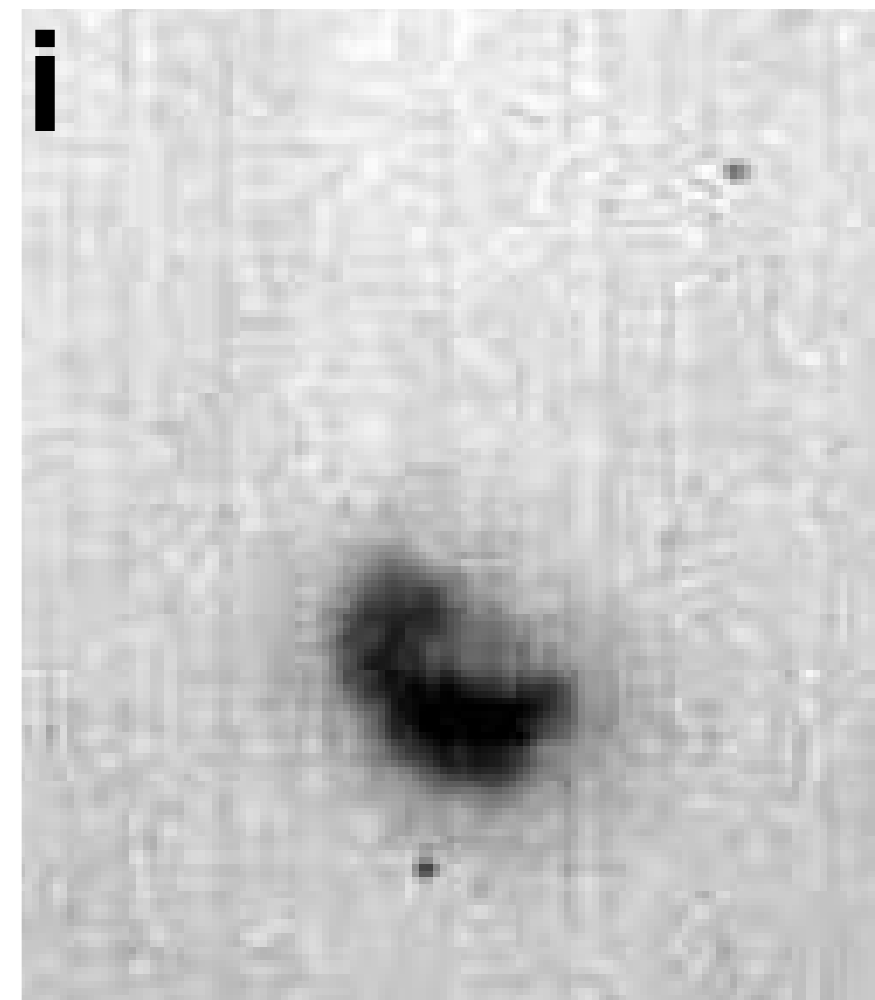} &
    \includegraphics[width=.10\textwidth]{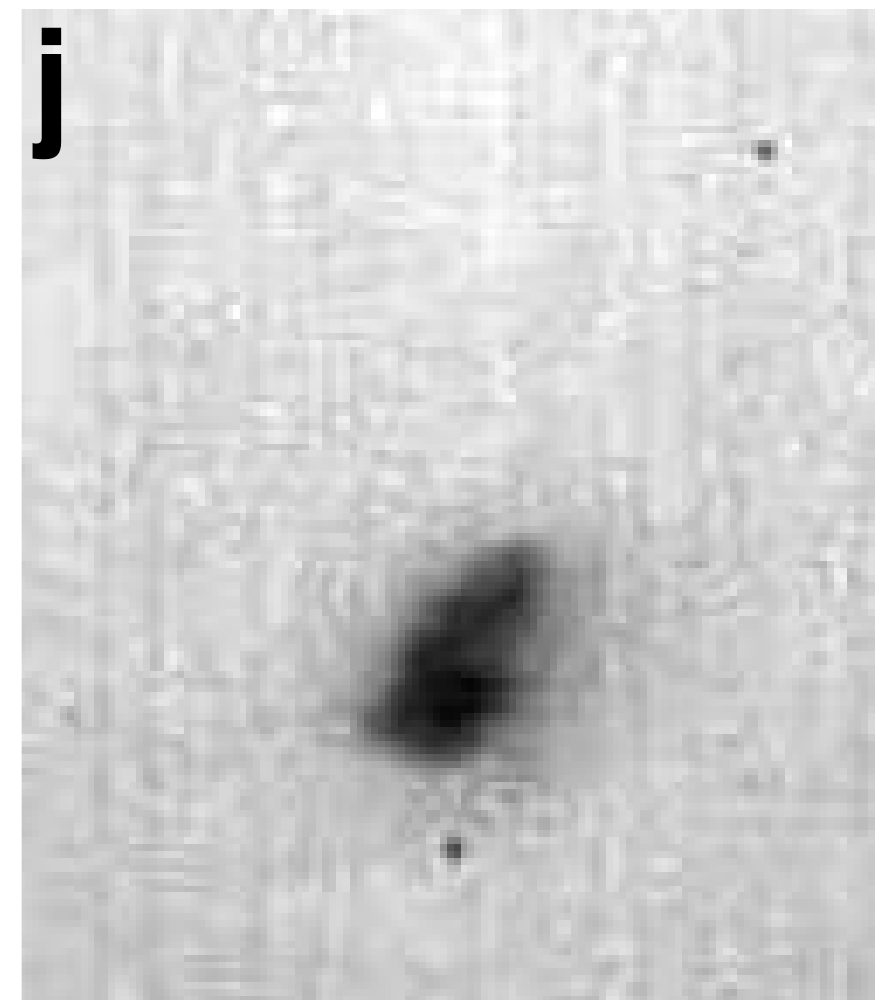} &
    \includegraphics[width=.10\textwidth]{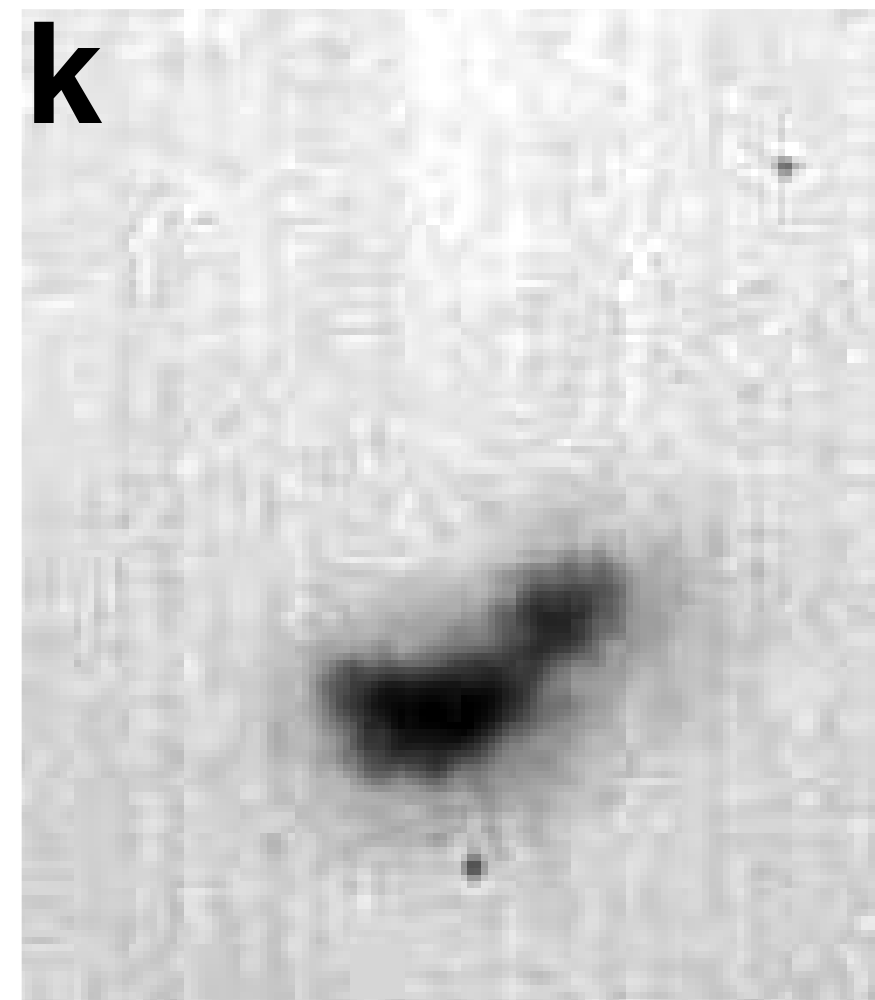} & 
\includegraphics[width=.10\textwidth]{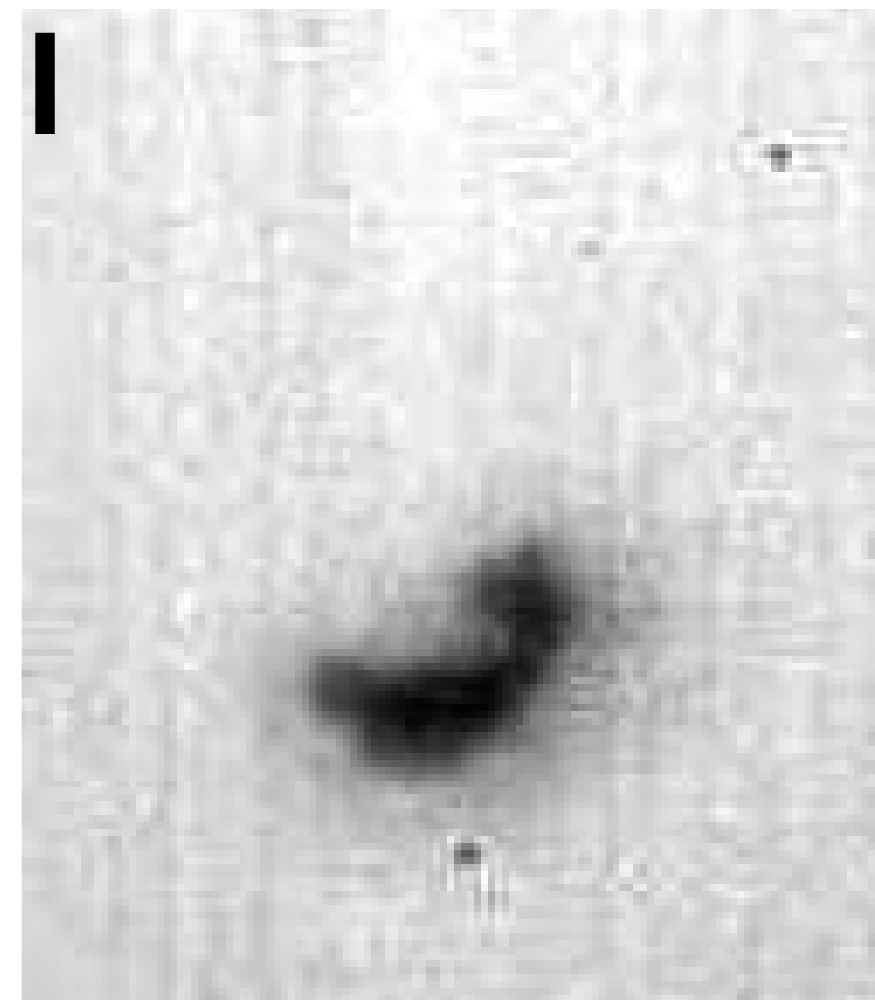}  \\
    \includegraphics[width=.10\textwidth]{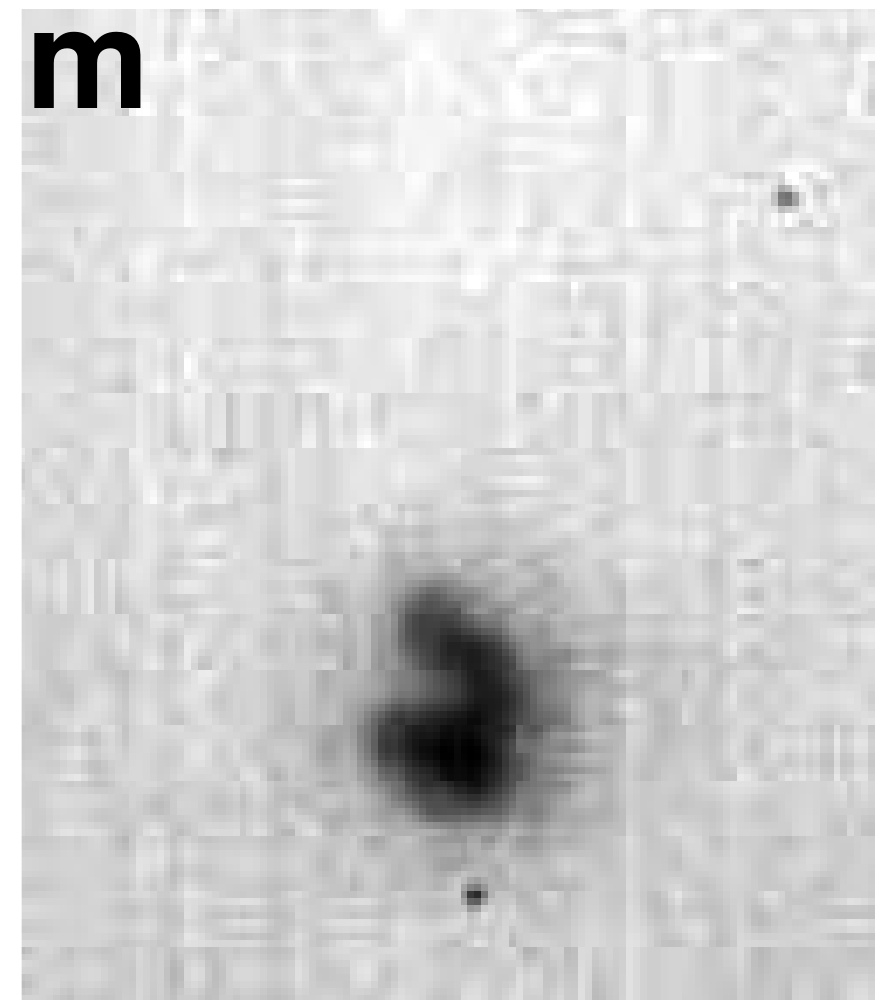} &
    \includegraphics[width=.10\textwidth]{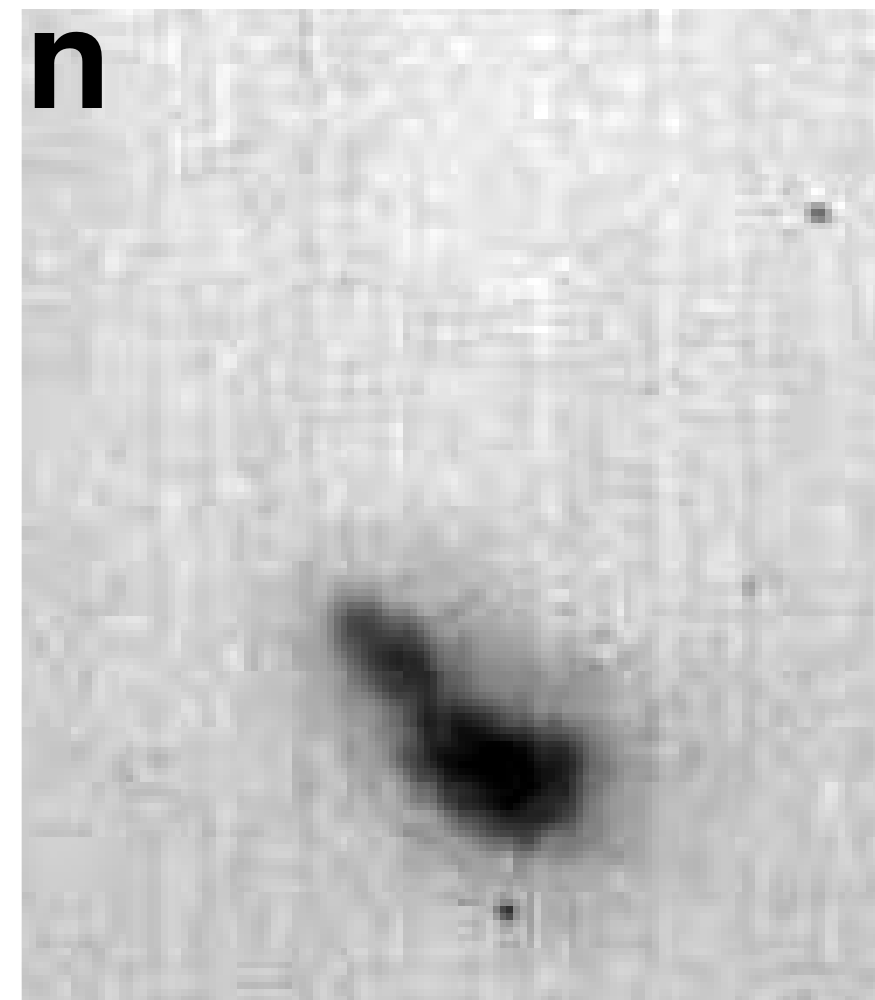} &
    \includegraphics[width=.10\textwidth]{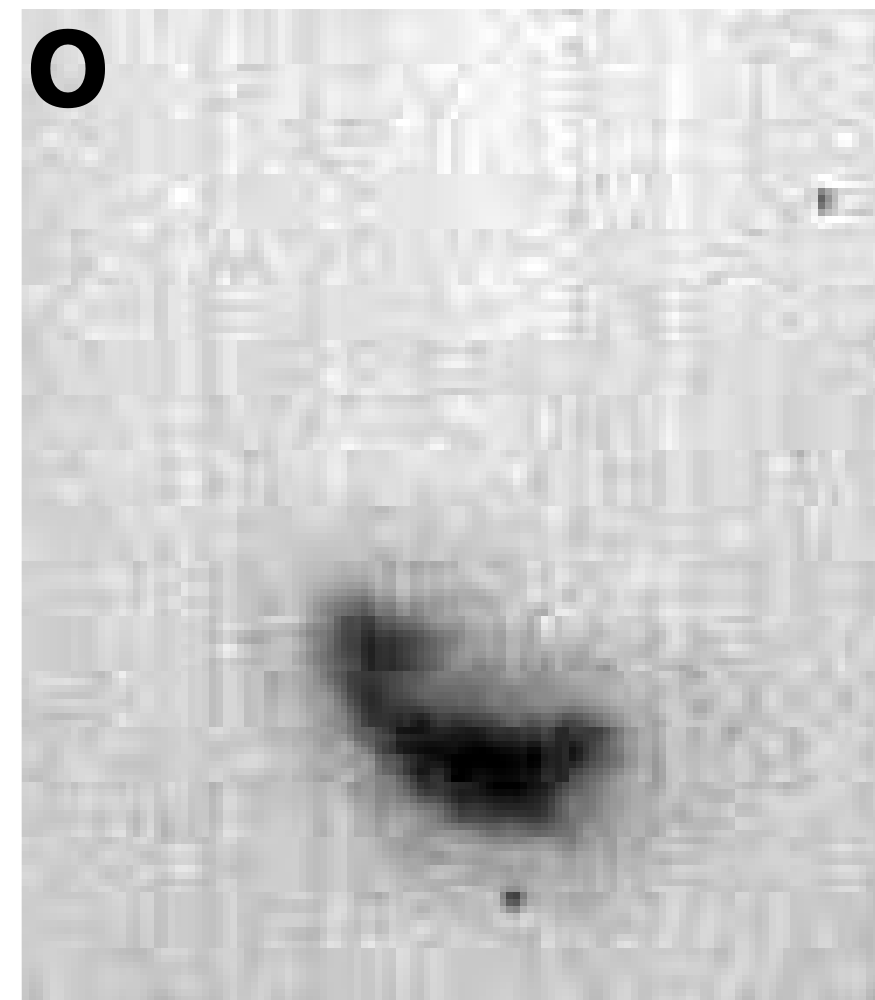} &
    \includegraphics[width=.095\textwidth]{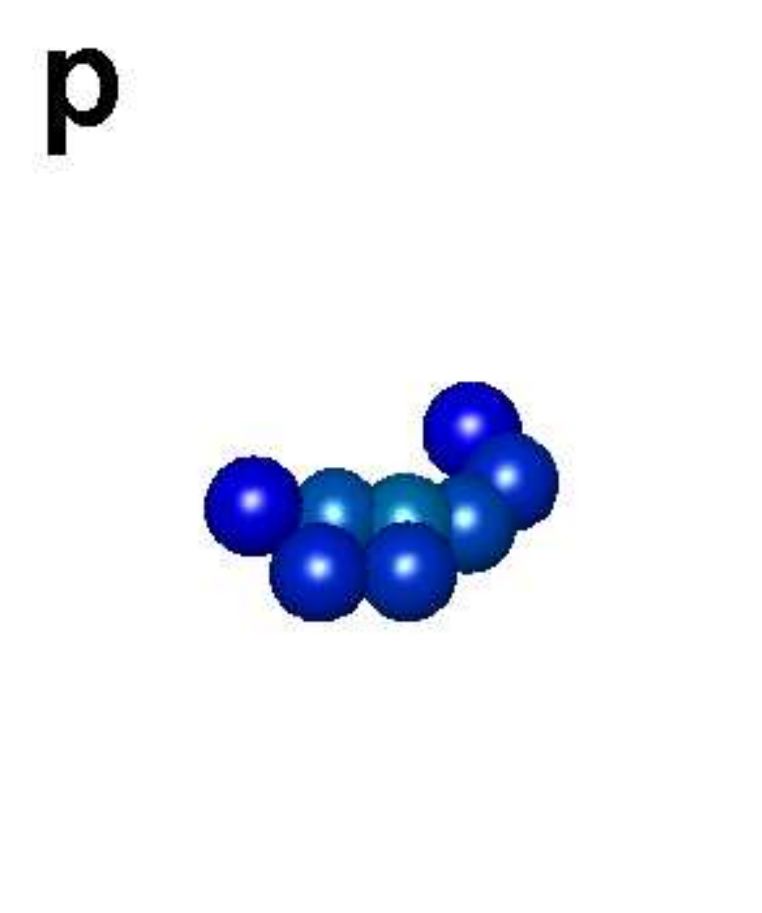}   \\
  \end{tabular}
  \caption{($a-g$) Sequence of images showing the complete rotation of cluster $P1$ from Fig. 1 consisting of eleven 8.93 $\mu$m gold-coated mf
 particles, $(\Delta{t}= .0167 s)$, and ($h$) its reconstructed 3D model. ($i-o$) $P2$ from Fig. 1 consisting of eight particles,  $(\Delta{t}= .0117 s)$, and ($p$) its 
reconstructed 3D model.}
\lb{fig2}

\end{figure}

In this work all complex plasma experiments are performed in a modified GEC (Gaseous Electronics Conference) RF reference cell.
 The vacuum chamber has two electrodes, a lower cylindrical electrode driven at a frequency of 13.56 MHz and a hollow cylindrical upper electrode, which is grounded.
  Single gold-coated melamine formaldehyde (mf) spheres with diameter of 8.94 $\mu$m,
 are introduced into the argon plasma from a shaker located above the upper electrode. 
 The dust particles are confined inside an open-ended glass box placed on the lower electrode. See \cite{9} for complete description of the experimental setup. The plasma is typically maintained at a RF
 peak-to-peak voltage of 80 V and a pressure of 500 mTorr, allowing the particles to form a stable cloud within the box. Using the same method as in \cite{10} for
 creating the aggregates, the particles are accelerated by triggering self-excited dust density waves by rapidly decreasing the pressure to 50 mTorr. After several minutes the pressure is returned
 to the initial level, and aggregates consisting of up to $\approx$20 monomers are observed.  
Aggregates are backlit using a 500 W flood lamp and imaged using a CMOS monochromatic high speed camera
 (1024 FASTCAM Photron) at 3000 fps.
 To ensure that aggregates form within the focal plane of the camera and lens assembly, the glass box is placed off-center on the lower electrode, closer to the window where the
 camera is mounted.  The recorded field of view spans an area of 717 $\mu$m $\times$ 829 $\mu$m with a resolution of 1.4 $\mu$m/pixel.

\begin{figure}
\hfill
\subfigure{\includegraphics[width=1.73cm]{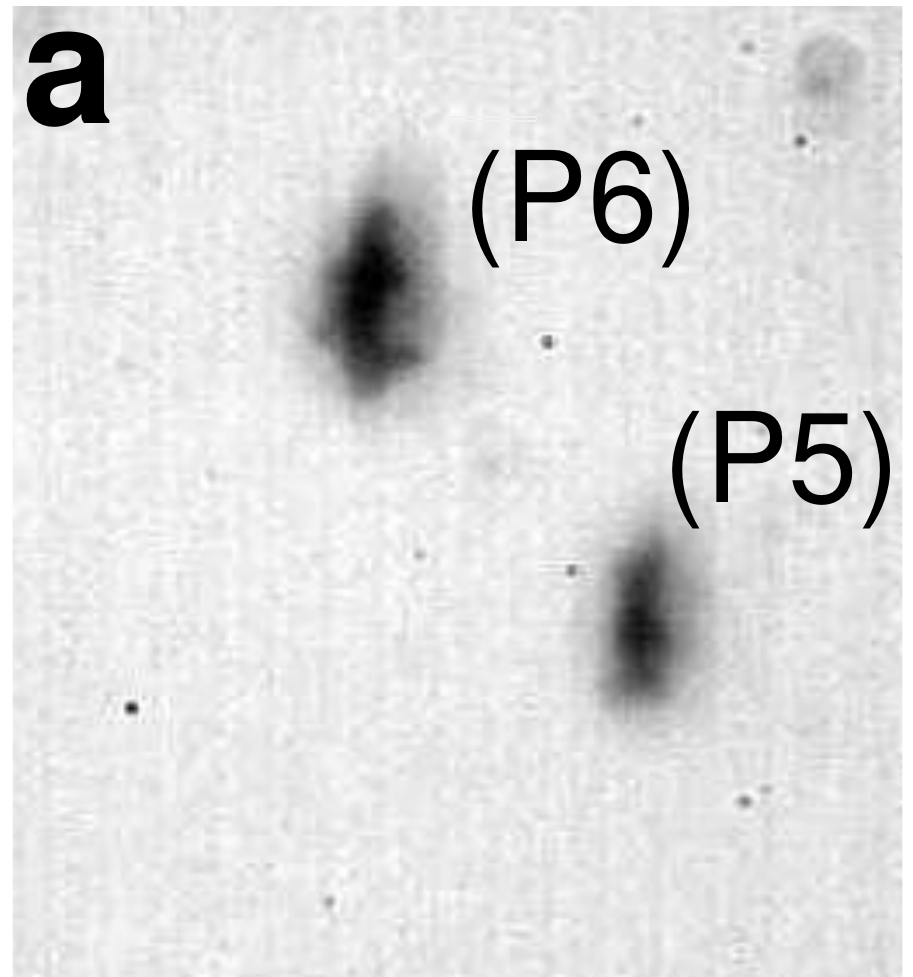}}
\hfill
\subfigure{\includegraphics[width=1.6cm]{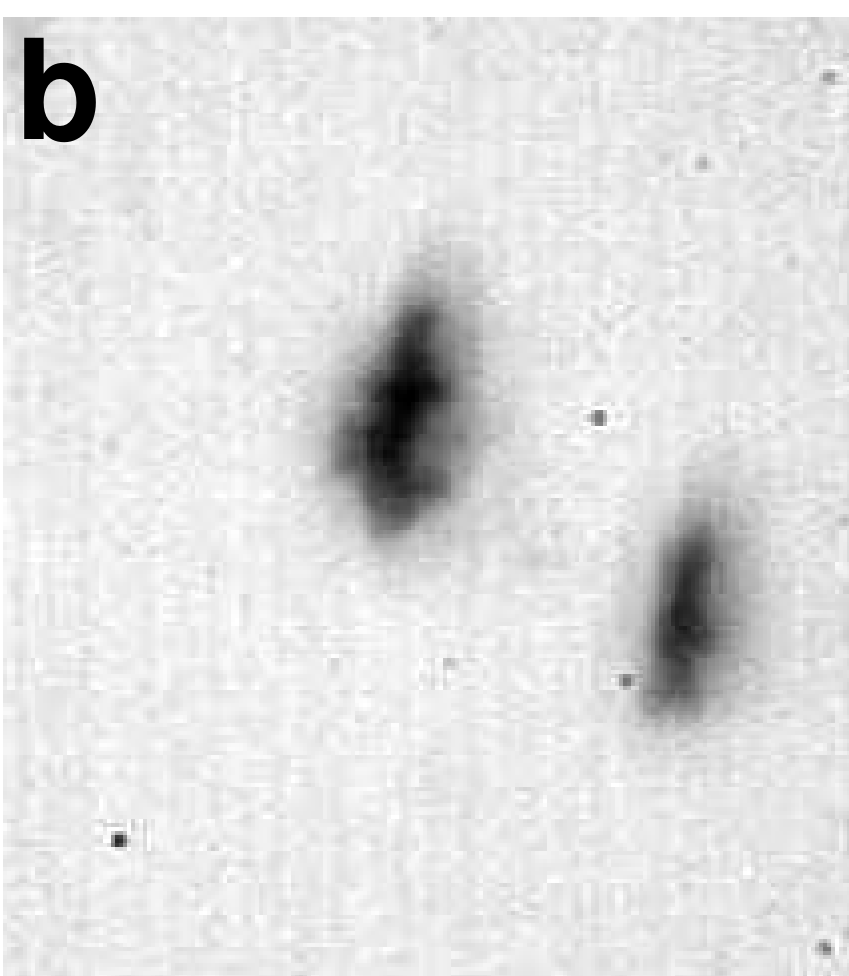}}
\hfill
\subfigure{\includegraphics[width=1.6cm]{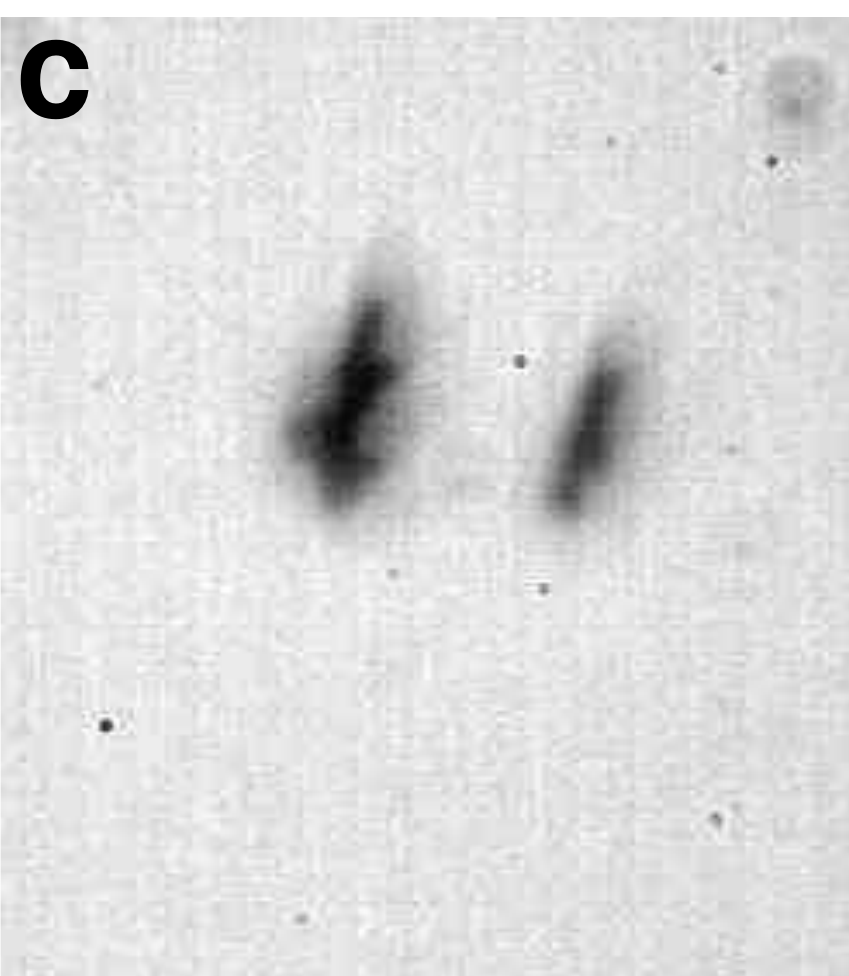}}
\hfill
\subfigure{\includegraphics[width=1.6cm]{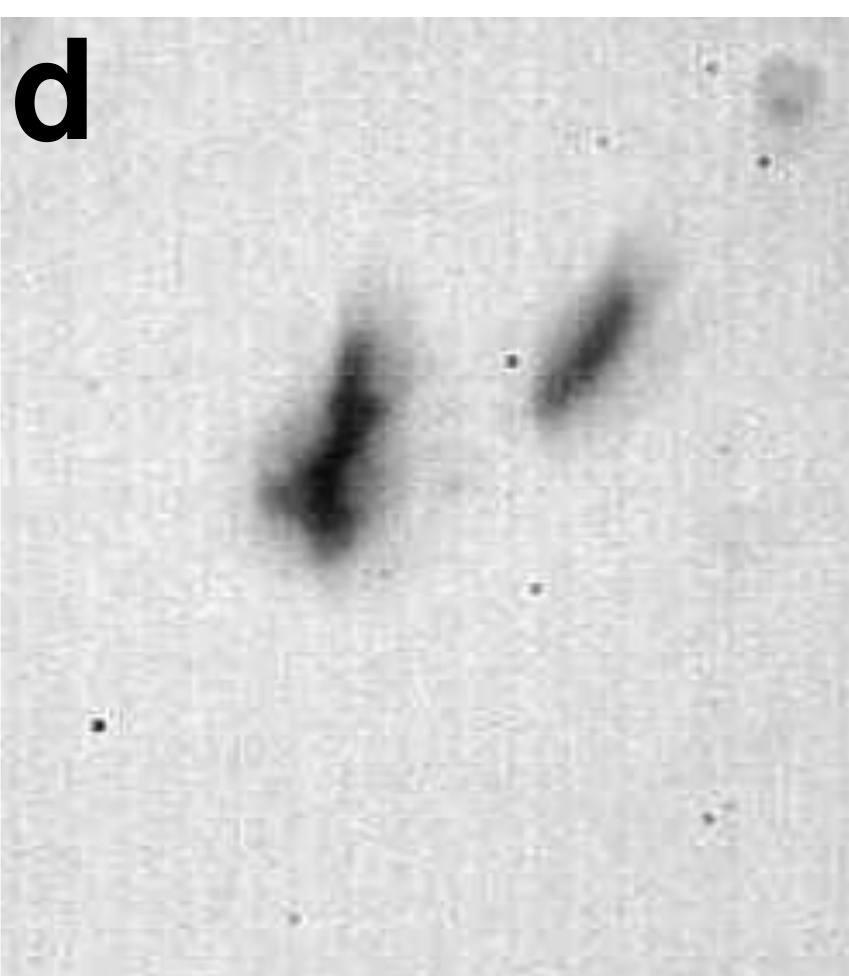}}
\hfill
\subfigure{\includegraphics[width=1.6cm]{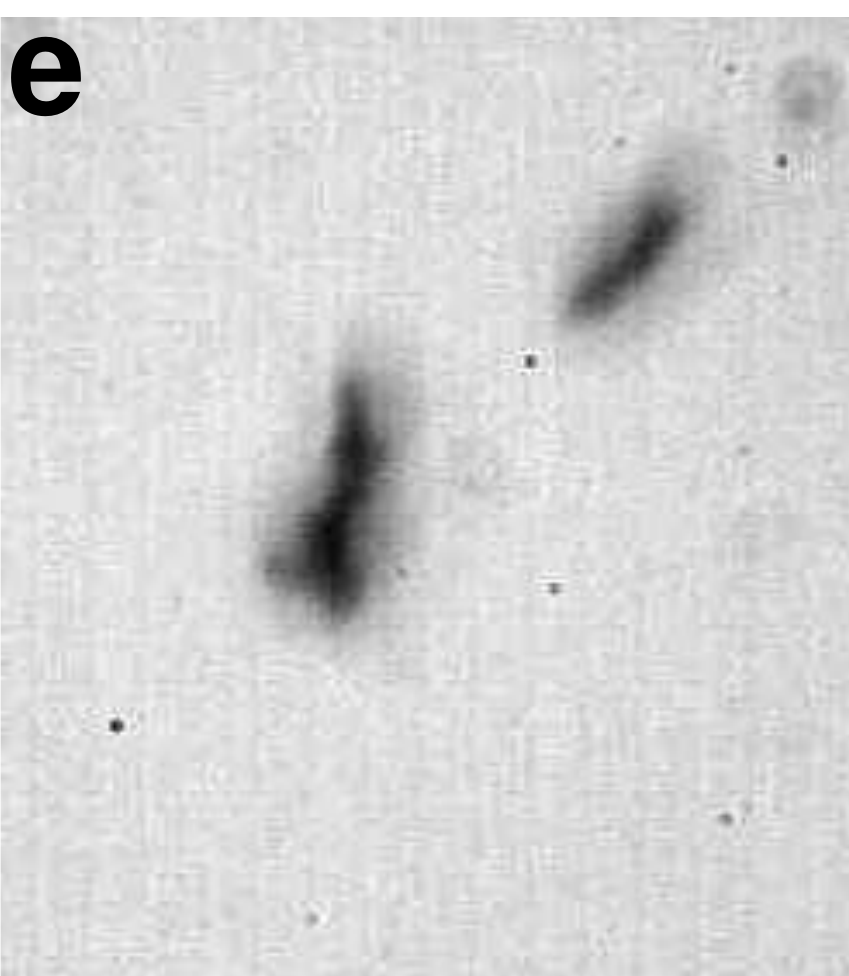}}
\hfil
\caption{A sequence of images showing a sample interaction between two charged dust particles ($P5$) and ($P6$). ($P5$) begins in the bottom right and ($P6$)  begins in the top left ($\Delta{t}$=0.001s)}.
\label{fig3}
\end{figure}

 In the experiment, observed aggregates tend to be linear and elongated.  Individual aggregates are often observed to rotate about the vertical z-axis, with a rotation period
 of approximately a few hundredths of a second. A representative sample of the aggregates observed is shown in Fig. \ref{fig1}. Aggregate rotation allows three-dimensional models to be
 reconstructed, as illustrated in Fig. \ref{fig2}. Interacting aggregate pairs exhibit rotation induced by their charge-dipole interactions as shown in Fig. \ref{fig3}.

\section{NUMERICAL MODEL}

Two numerical codes are used to model the charging and interaction of the aggregates. The electric charge and dipole moment are calculated using $OML_{-} LOS$, which is based
 on $OML$ (orbital motion limited) theory 
modified to determine the open lines of sight ($LOS$) to points on the surface of the aggregate \cite{13}. Electrons and ions in the plasma are incident to the surface
 along the open LOS, yielding a charge distribution over the aggregate surface. Assuming an electron temperature, $T_e = 23400$ K ($\approx$ 2 eV), ion temperature, $T_i = 298$ K,  and $n_e = n_i = 1 \times10^{14}$ m$^{-3}$ where $n_e$ is the 
number density of electrons and $n_i$ is number density of ions,
 the total charge and dipole moment on each aggregate is calculated. Note that the values derived from ($OML_{-}LOS$) code are very sensitive to the number of 
monomers and their exact arrangements used to build up the 3D structure of the aggregates. The interaction between two charged
 grains including  rotations induced by torques due to the 
charge dipole moments, is modeled using the $Aggregate _{-}Builder$ code \cite{14}.

\section{ANALYSIS}

In the experiment, the rate of change of the velocity $\vec{v}$ for particles with mass $m$ and charge $q$, was calculated using the equation
 of motion including gravity, $\vec{F}_G$, gas drag, $ \vec{F}_D$, and the electrostatic force, $ \vec{F}_E$, acting on the particle,

\begin{equation*}
\lb{1}
m\frac{d\vec{v}}{dt}=\vec{F}_G + \vec{F}_D + \vec{F}_E. 
\tag{1}
\end{equation*}

For weakly ionized plasmas, 
we consider only neutral gas drag since it dominates the ion drag \cite{12}. The neutral drag force is given by the Epstein formula \cite{12aa}

\begin{equation*}
\lb{2}
 \vec{F}_D = - \beta m \vec{v}
\tag{2}
\end{equation*}
where $\beta$ is the gas drag coefficient, which was experimentally determined to be $\beta$ = 9.8 s$^{-1}$ for small mf aggregates in argon plasma at 500 mTorr using the same 
method as in \cite{12a}.

\begin{figure}
\hfill
\subfigure[ Raw image]{\includegraphics[width=2.58cm]{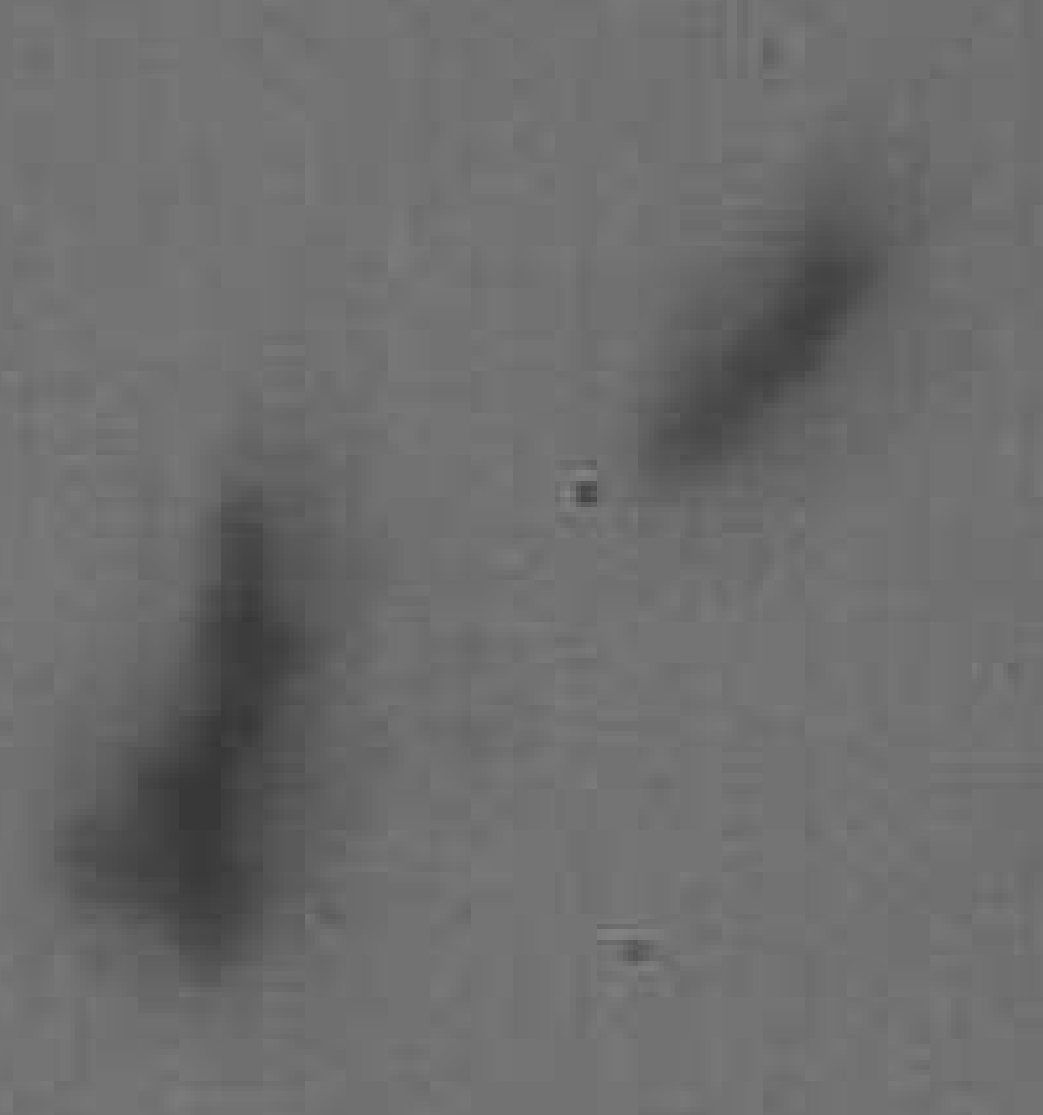}}
\hfill
\subfigure[ Binary image]{\includegraphics[width=2.6cm]{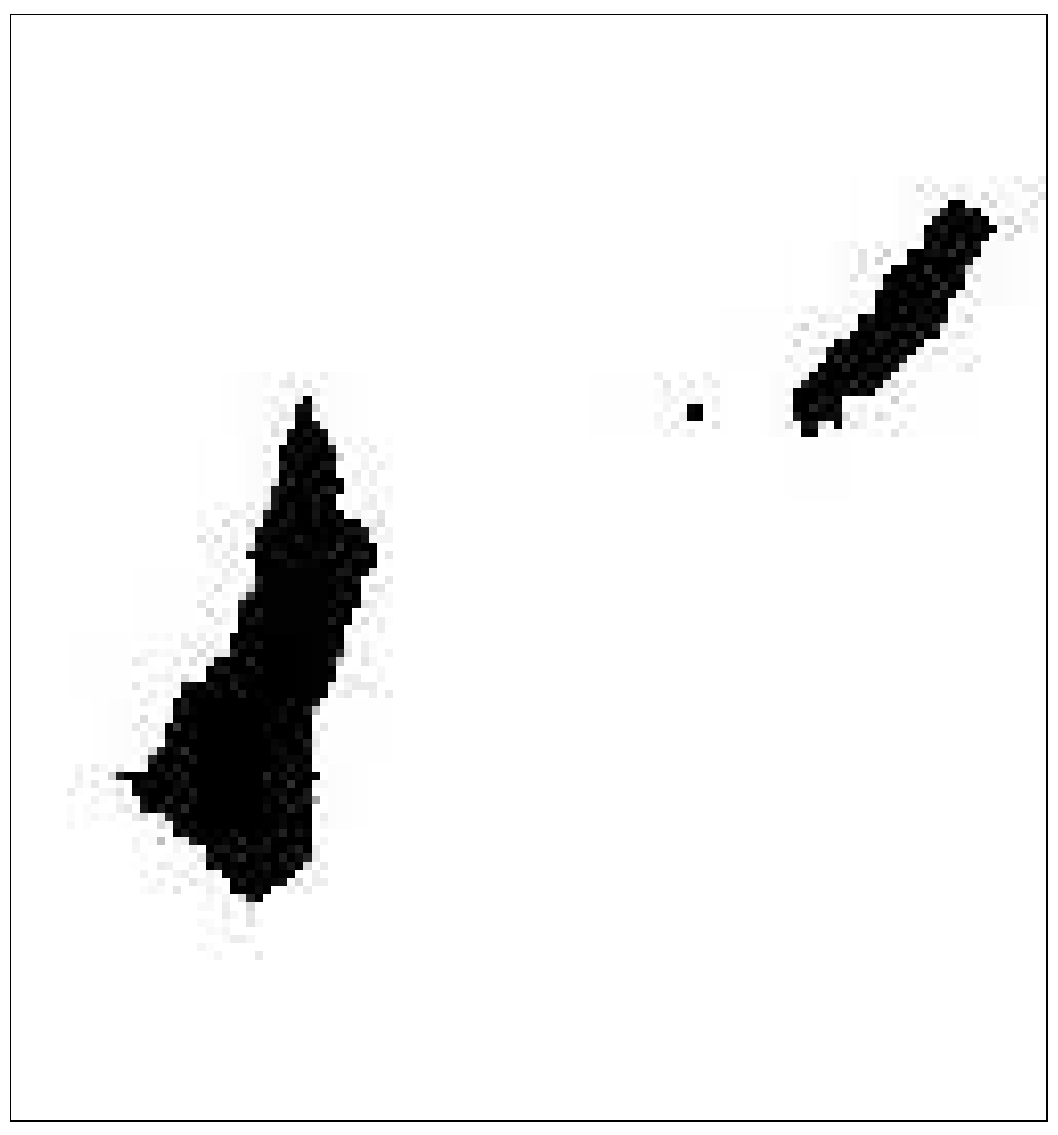}}
\hfill
\subfigure[ Analyzed image]{\includegraphics[width=2.6cm]{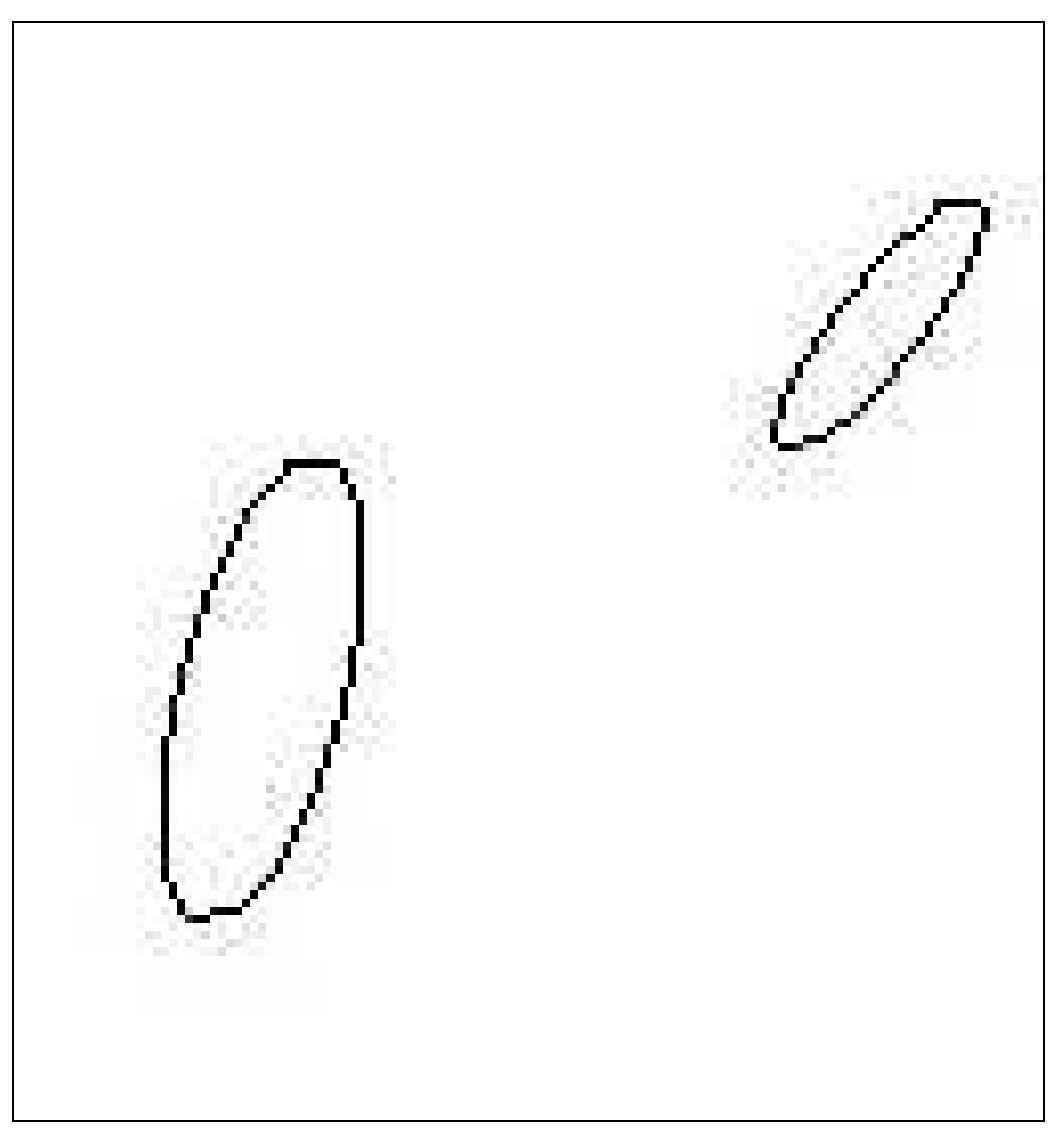}}
\hfill
\caption{Illustration of image processing. (a) A raw image loaded into ImageJ. (b) Binary image after setting a pixel value
 threshold in the program for the image shown in (a). (c) Image analysis by fitting ellipses snugly around the detected aggregates in (b) through ``Analyze Particle'' function. }
\label{fig4}
\end{figure}

The electrostatic force arises from the interaction of dust particle with charge $q$ with the electric field from other dust grains, $\vec{E}_{int}$,
 the existing vertical electric field in the sheath above the lower electrode, $\vec{E}_z$, and a confining electric field in the horizontal direction arising from walls of the box, $\vec{E}_{r}$,

\begin{equation*}
\lb{3}
 \vec{F}_E = q (\vec{E}_{int}+\vec {E}_z + \vec{E}_{r})
\tag{3}
\end{equation*}

 The electric field produced by a charged particle can be expanded as
\begin{equation*}
\lb{4}
\vec{E} = \frac{1}{4 {\pi}{ \epsilon} _{0}}   \{   \frac{ q \vec{r}}{r^3} + \frac{3\vec{r}(\vec{p} \cdot \vec{r} - \vec{p}) }{r^5} + \cdots   \}
\tag{4}
\end{equation*}
where $\vec{p}$ is the electric dipole moment and $\vec{r}$ is the vector displacement from the particle's center of mass. The electric field is dominated by the leading non-vanishing term in
$\frac{ \vec{r}}{r^3}$, all the higher-order terms being negligible for calculating forces acting on the particles. It is important to note however that the 
electric dipole moment of the particles plays a very important role in determining the orientation of interacting particles.

 The rotation of a particle is 
driven by the total torque acting on the dust particle. The angular acceleration $\alpha$
 of the particle is given by

\begin{equation*}
\lb{5}
 \frac{d {\vec{\L}}}{dt}  = I \vec{\alpha}     = \vec{\tau}_E +\vec{\tau}_D
\tag{5}
\end{equation*}
where $\vec{\L}$ is angular momentum, $I$ is the moment of inertia of the aggregate, $\vec{\tau}_{E} = \vec{p} \times \vec{E}$ is the torque due to the electric field including $\vec{E}_{int}$, $\vec{E}_{z}$ and $\vec{E}_{r}$, and
 $\vec{\tau}_D =\sum_i \vec{r_i} \times \vec{F}_D$ is the torque due to the drag force, with $\vec{r}_i$ being the perpendicular distance 
from the center of each monomer to the vertical axis about which the aggregate is rotating.

\section{RESULTS}

For rotating aggregates, a sequence of $500$ image frames was analyzed using ImageJ \cite{12b}. A number of
 aggregates, almost fixed in place, spinning about the vertical z-axis with a nearly constant angular velocity 
were observed (Fig. \ref{fig1}). Two of the particles, $P1$ and $P2$, which are in focus were examined using the 3D models shown in Fig. 2(h) and 2(p).
 The total charge on these rotating aggregates was estimated numerically using the $OML_{-}LOS$ 
code to be $q_{1} = -0.82 \times 10^{-14}$ C and $q_{2} = -0.59 \times 10^{-14}$ C, respectively.
  Since these aggregates show little horizontal translational motion, the gravitational
 force acting on the particles is assumed to be balanced by the electrostatic force due to the vertical electric field. 
The Debye length of the plasma is estimated to be about $0.1$ $mm$ using the values of $T_e$, $T_i$, $n_e$ and $n_i$ provided by $OML_{-}LOS$. The minimum distance
 between the rotating aggregates in the 2D images is measured to be $\sim 0.3$ mm which is larger than the calculated Debye length.
 Therefore it is reasonable to assume that $\vec{E}_{int}$ is negligible between the particles. Using this fact, the magnitude of the vertical electric field, corresponding to the position of each aggregate is
 estimated to be $E_{z1}= 0.80\times10^4$ V/m and $E_{z2}=0.81\times10^4$ V/m. The resulting electric field gradient is consistent with the results reported in \cite{12baa}.

\begin{figure}
\centering
{\includegraphics[width=.488\textwidth]{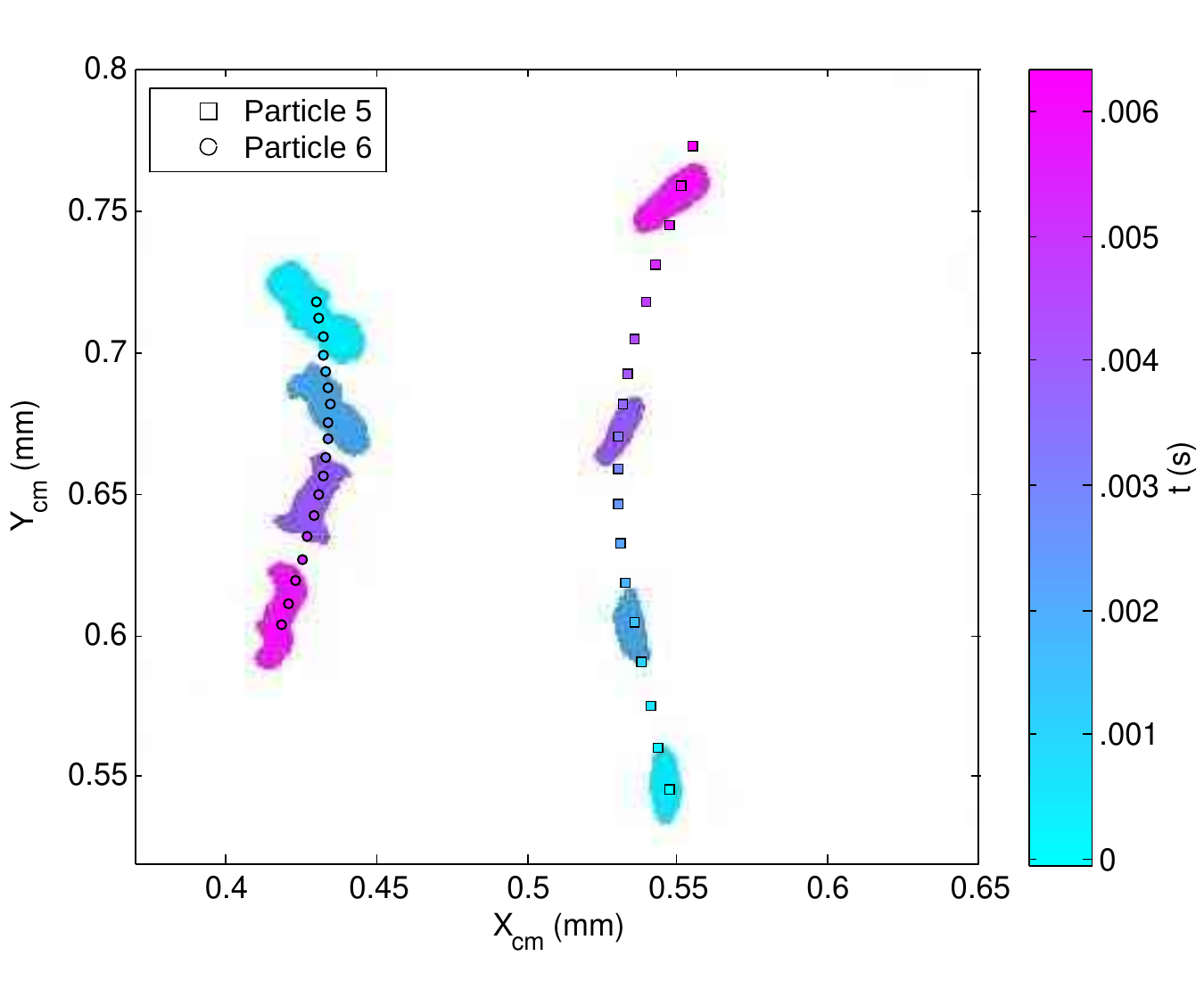}}
\caption{Time evolution of the $X$- and $Y$-components of center of mass position for two interacting particles, P5 and P6,
superposed on four frames from Fig. 3, with color indicating the time.}
\label{fig5}

\end{figure}

Since the electrostatic force acts at the ``center of charge'' while gravity acts at the center of mass, the two forces exert an aligning torque, with any rotation about a horizontal axis damped by gas drag.  
Thus, rotation is about the vertical axis only. Using Eq. (\ref{5}), the magnitude of the electric dipole moment can be calculated in the horizontal plane. The electric field which is responsible
 for this rotation is in the horizontal direction, $\vec{E}_r$, which is assumed to be the confining electric field created by the negative potential on the walls of the box. An
 estimation of the magnitude of this horizontal electric field using the same experimental set up is given in \cite{12bb} and taken to be $\mid\vec{E}_{r}\mid = 100$ V/m.
 As the observed particles are rotating with a constant angular velocity, the total torque applied on each particle due to this external electric field is 
balanced by the torque due to the drag force. Using this fact, the magnitude
 of the electric dipole moment in the horizontal plane is calculated to be $\mid \vec{p}_{1r}\mid =3.1\times10^{-19}$ Cm and  $\mid \vec{p}_{2r}\mid =0.31\times10^{-19}$ Cm. These values are in very good agreement
 with the values predicted by $OML_{-}LOS$, $\mid \vec{p}_{1r}\mid =3.1\times10^{-19}$ Cm and  $\mid \vec{p}_{2r}\mid =0.29\times10^{-19}$ Cm. It is interesting to note that the value of the horizontal electric dipole moment for $P4$, which is 
observed to have no rotation during the measured time frame, is predicted to be $\mid \vec{p}_{1r}\mid =0.09\times10^{-19}$ Cm by $OML_{-}LOS$. The very small value of the horizontal electric dipole moment could be one reason for the lack of rotation.

\begin{figure}
\includegraphics[width=.508\textwidth]{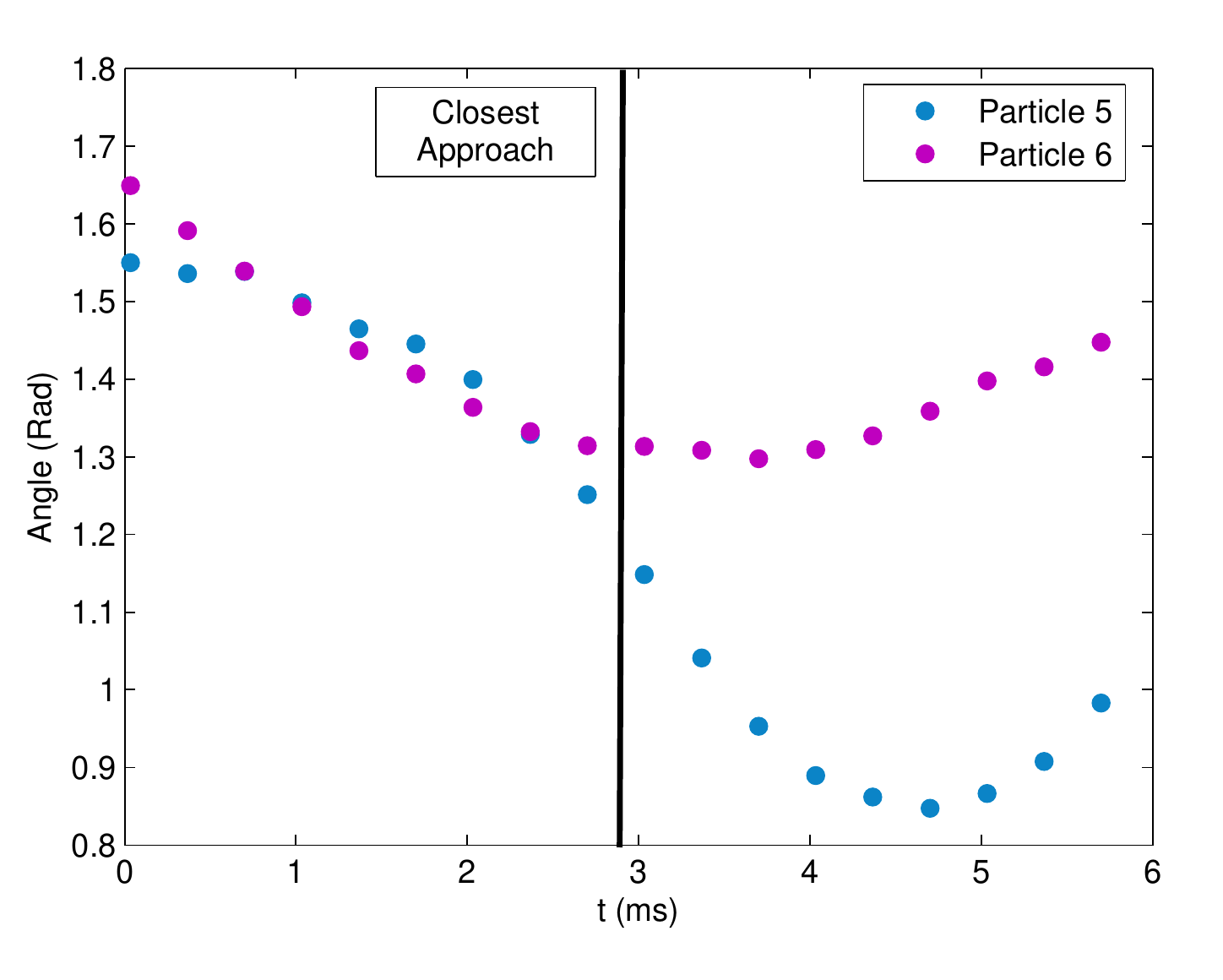}
\caption{  Orientation angle verses time for P5 and P6 with the solid line showing the closest approach. }
\label{fig6}

\end{figure}

\begin{figure}[htb]
\centering
  \begin{tabular}{@{}cccc@{}}
\hfill
\subfigure[]{\includegraphics[width=.242\textwidth]{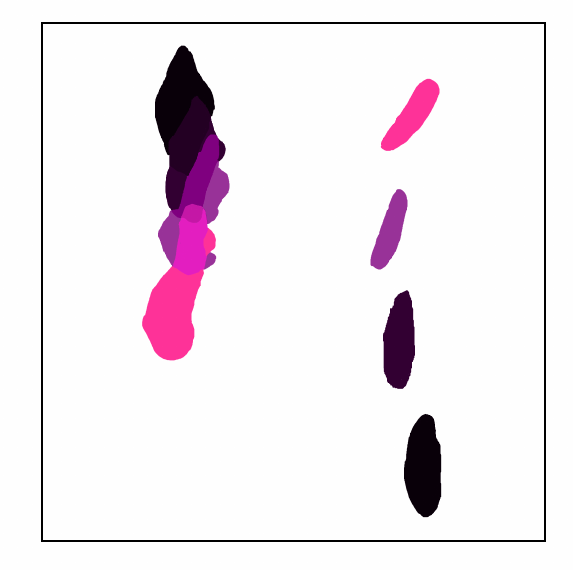}}
\hfill
\subfigure[]{\includegraphics[width=.242\textwidth]{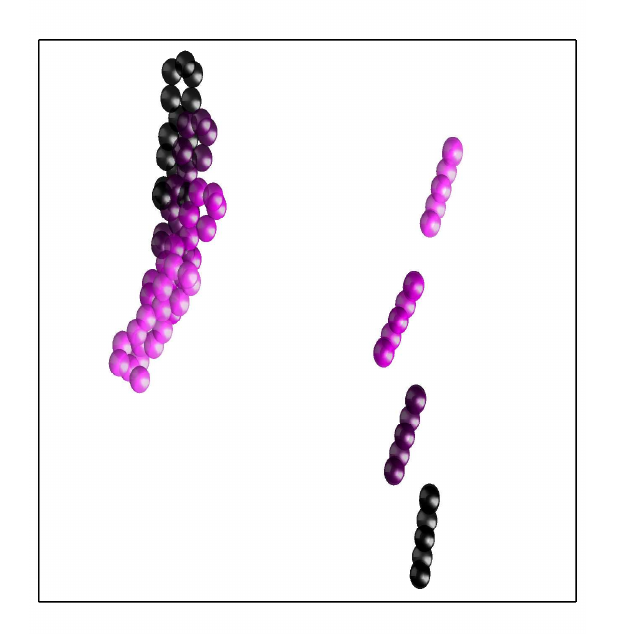}}\\
\hfill
   
  \end{tabular}
  \caption{(a) A superposition of four frames showing the interactions between two aggregates from experimental data with similar color of aggregates in each frame with (b) a superposition of four
 frames from $Aggregate_{-}Builder$ code with similar color of aggregates in each frame, showing a similar interaction of two aggregates. Not all monomers are visible.}

\label{fig7}
\end{figure}

 For the interacting aggregate pairs, a 30-frame sequence of images, showing a short interaction between two charged dust particles, $P5$ and $P6$,  
 was analyzed using ImageJ. As shown in  Fig. \ref{fig3}, $P5$ enters the frame from the lower right and $P6$ enters the frame
 from the upper left. As they approach one another, their paths are deflected. $P5$'s rotation about an axis perpendicular to the image plane is considerable.
 A pixel value threshold was set in the program to convert the gray-scale image into a binary image (Fig. 4(b)). 
The ``Analyze Particles'' function in ImageJ was then used to fit ellipses snugly around the particles detected as shown in Fig. \ref{fig4}(c) to determine the coordinates and orientations of all the
 particles in each frame. Fig. \ref{fig5} shows the center of mass positions
 $X_{cm}$ and $Y_{cm}$ for both particles as they change over time overlaid on four frames from the movie, with their orientation as a function of time shown in Fig. 6. A polynomial function was fit to 
 the X- and Y-position data for each particle, with velocity and acceleration of the particles derived by differentiating this equation \cite{12bbb}.

The two vector-component equations from Eq. (\ref{1}) for the two particles at the point of closest approach reduce to
 three independent equations with three unknown variables: $ q_5$, $q_6$ and $\vec{E}_z$, where $ q_5$ is the net charge on $P5$ and $ q_6$ is the net charge on $P6$.
 Solving these equations yields $ q_5= -0.43\times 10^{-14}$ $C$, $q_6 = -0.72\times 10^{-14}$ $C$ and $ \mid \vec{E}_z\mid = 1.3\times10^{4}$ $V/m$  in the downward
 direction.

The electric dipole moments of the dust grains are then calculated from the particles' changing orientations during the interaction.
The angular acceleration of each particle 
was calculated from the orientation, as shown in Fig. \ref{fig6} using the values of charges and external electric field found in the previous step, and then substituting these 
values into Eq. (\ref{5}). The electric dipole moments for each of the dust grains were found to be 
$\mid \vec{p_5}\mid =0.25\times10^{-20}$ Cm and $\mid \vec{p_{6}}\mid =2.8\times10^{-20}$ Cm.

 Using reconstructed 3D models of P5 consisting of eight monomers and P6 consisting of sixteen monomers in $OML_{-}LOS$, charges and electric dipole moments
 of aggregates were calculated as  $q_5 = -0.58\times 10^{-14}$ C and $q_2 = -0.83\times 10^{-14}$ C with corresponding components of dipole moments as $\mid \vec{p_{5}}\mid =0.24\times 10^{-20}$ Cm and 
$\mid \vec{p_{6}}\mid = 2.7\times 10^{-20}$ Cm. These values are in very good agreement with the values derived from experiment. The minor differences seen are assumed to arise from the sensitivity of the $OML_{-}LOS$
 code to the number of monomers and their exact orientation in the constructed 3D aggregates.

The accuracy of the derived values are tested by modeling the dynamics of the interaction in $Aggregate-Builder$. Fig. \ref{fig7} compares the observed and the modeled
 interaction where charges and electric dipoles are those predicted by $OML_- LOS$. 
Note that the trajectories of particles are sensitive to the interacting particles' charges and electric dipole moments as well as their initial positions, velocities and orientations. 
The forces arising from the external electric fields are quite small compared to the Coulomb interaction between the aggregates and have a small impact on the particles' trajectories.

\section{CONCLUSIONS}

 Aggregates were formed from gold coated 
spherical melamine-formaldehyde monomers in a rf argon discharge plasma. The electrostatic charges and dipole moments of these aggregates were determined
 through analysis of the extracted particle trajectories and rotations and compared to the output of the numerical models. The excellent agreement between experiment and simulation 
validates the choice of underlying assumptions. For the numerical models, these include the simplifying assumptions that ions and electrons 
trajectories only impact the aggregate surface along open lines of sight and stick at the point 
of impact and that rotations of charged grains are induced by charge-dipole interactions. In deriving values from experiment, assumptions made were neglecting the ion drag force, considering a coulomb interaction 
between charged particles in order to calculate the force between them, and ignoring the gradient in $E_z$ and $E_r$
 during the time of interaction for interacting pair particles. To the best of our knowledge, this is the first study designed to investigate dust aggregates with irregular shapes in 
order to evaluate the electrostatic charge and dipole moment. In previous works aggregates were 
simplified as spherical grains and many neglected the treatment of their electrostatic dipole moments. These measurements of the electrostatic charges and dipole
 moments are of importance in understanding the aggregation process, in dynamic simulations studying dust agglomeration where the 
charge and electric dipole moment play a very important role and determining the evolution of the grain-size distribution.

\section*{ACKNOWLEDGMENTS}

This work is supported by the National Science Foundation under Grant No. 0847127.


\end{document}